\documentclass[journal]{IEEEtran}
\IEEEoverridecommandlockouts

\usepackage{amsmath}
\usepackage{amsfonts}
\usepackage{amssymb}
\usepackage[ruled,vlined]{algorithm2e}
\usepackage{caption}
\usepackage{mathtools}  
\usepackage{amsthm}
\usepackage[]{algorithm2e}
\usepackage{booktabs}
\usepackage{subfigure}
\usepackage{tabulary}
\usepackage{footnote}
\usepackage[export]{adjustbox}
\usepackage{booktabs}
\usepackage[justification=centering]{caption}
\usepackage{comment}
\newtheorem{remark}{Remark}
\usepackage{cases}
\usepackage{graphicx}
\usepackage{cite}
\usepackage{multicol}
\usepackage{xcolor}
\usepackage{graphicx}
\graphicspath{{./}{figures/}}
\newcommand\numeq[1]%
  {\stackrel{\scriptscriptstyle(\mkern-1.5mu#1\mkern-1.5mu)}{=}}
\usepackage{stfloats}
\usepackage{cuted}
\usepackage{lipsum}  

\usepackage{subfigure}
\usepackage{dsfont}
\usepackage{amsmath,amssymb}

\newtheorem{lemma}{Lemma}

\title{Efficient PHY Layer Abstraction under Imperfect Channel Estimation}

\author{\text{Liu Cao, Lyutianyang Zhang, Sian Jin, and Sumit Roy, \IEEEmembership{Fellow,~IEEE}}

\thanks{Liu Cao, Lyutianyang Zhang, and Sumit Roy are with the Department of Electrical and Computer Engineering, University of Washington, Seattle, WA 98195 USA (e-mail: liucao@uw.edu; lyutiz@uw.edu; sroy@uw.edu). Sian Jin is with the Department of Electrical and Computer Engineering, Princeton University, Princeton, NJ 08544 USA (sj2434@princeton.edu) (\emph{Corresponding author: Liu Cao}.)}}%

\begin{document}

\maketitle
\thispagestyle{empty}
\begin{abstract}
As most existing work investigate the PHY layer abstraction under an assumption of perfect channel estimation, it may become unreliable if there exists channel estimation error in a real communication system. This letter improves an efficient PHY layer method, EESM-log-SGN PHY layer abstraction, by considering the presence of channel estimation error. We develop two methods for implementing the EESM-log-SGN PHY abstraction under imperfect channel estimation. We show that the effective SINR is not impacted by the channel estimation error under multiple-input and single-output (MISO)/single-input and single-output (SISO) configuration, which is also verified by the full PHY simulation. The developed methods are then validated under different orthogonal frequency division multiplexing (OFDM) scenarios.

\end{abstract}


\section{Introduction}
\label{sec:intro}
 \IEEEPARstart{F}{ull} PHY (link) simulations, which typically run on link simulators, quantify the network performance in Wi-Fi (IEEE 802.11 WLANs) systems from the packet-level performance metrics. However, running a full PHY simulation that involves generating channel realizations and transceiver signal processing is impractical within a network simulator \cite{jin2021efficient}. In addition, with increasing node density and complexity in the PHY layer, associated computational complexity is also a major concern for network simulators \cite{Vienna}. To cope with such issues, {\em PHY layer abstraction} \footnote{The PHY layer abstraction in a network simulator is the packet error model that produces a decision on whether the packet is successfully received or not based on the PHY layer setup.} is used to generate accurate link performance in a network simulator. In particular, a novel EESM-log-SGN PHY layer abstraction method \footnote{EESM-log-SGN
PHY layer abstraction \cite{jin2021efficient} bypasses the matrix calculation steps required by the traditional PHY layer abstraction methods, and directly models
effective SINR by using a 4-parameter distribution called log-SGN distribution.} achieves good accuracy in modeling the PHY layer performance, meanwhile, its runtime is insensitive to system dimensionality change and is much reduced as compared to the traditional PHY layer abstraction methods.

To achieve PHY layer abstraction of current Wi-Fi systems which operate over wideband frequency-selective fading channels, link-to-system (L2S) mapping has been widely adopted \cite{RBIR11ax,11axEva,phyAbsMatlab,phyAbsCompar,brueninghaus2005link, wan2006fading}. L2S mapping function translates the post processing Signal-to-Interference-plus-Noise-Ratio (SINR) matrix into a single scalar metric called the {\em effective SINR} \cite{heath}. Abstracting the post processing SINRs over all subcarriers and spatial streams using a single valued effective SINR significantly simplifies the L2S interface \cite{wan2006fading, eesmVsRbir}.
The effective SINR is
a convenient metric to describe the packet-level performance in a network simulator, at the cost of losing detail for each
spatial stream (symbol-level performance), and fully align with the packet-level performance obtained from the full PHY simulations.

In any communication system, channel estimation is implemented by the receiver. However, the estimated channel is inherently noisy in practical systems, which incurs channel estimation error. Such channel estimation error may severely impact the post processing SINR,  further impacting the effective SINR which characterizes the results related to PHY layer abstraction. To the best knowledge of the author of this letter, the channel estimation error was hardly investigated in PHY layer abstraction as the channel estimation was assumed to be noise free in previous work. As a result, one key issue arises that the PHY layer abstraction executed under perfect channel estimation may become no longer reliable since any real communication system cannot perform perfect channel estimation.

Motivated by the aforementioned issues, it is necessary to investigate the PHY layer abstraction impacted by the channel estimation error. This letter develops the EESM-log-SGN PHY layer abstraction in the presence of the channel estimation error, which accords with the current PHY layer abstraction implementation as well as the MIMO architecture of typical Wi-Fi systems. Meanwhile, an analytical model is proposed to capture the relation between the channel estimation error and the effective SINR. The analytical effective SINRs are further utilized to develop the EESM-log-SGN PHY layer abstraction.

\section{System Architecture}
\label{sec:sys_arc}
\begin{figure}[h]
    \centering
    \includegraphics[width=.4\textwidth]{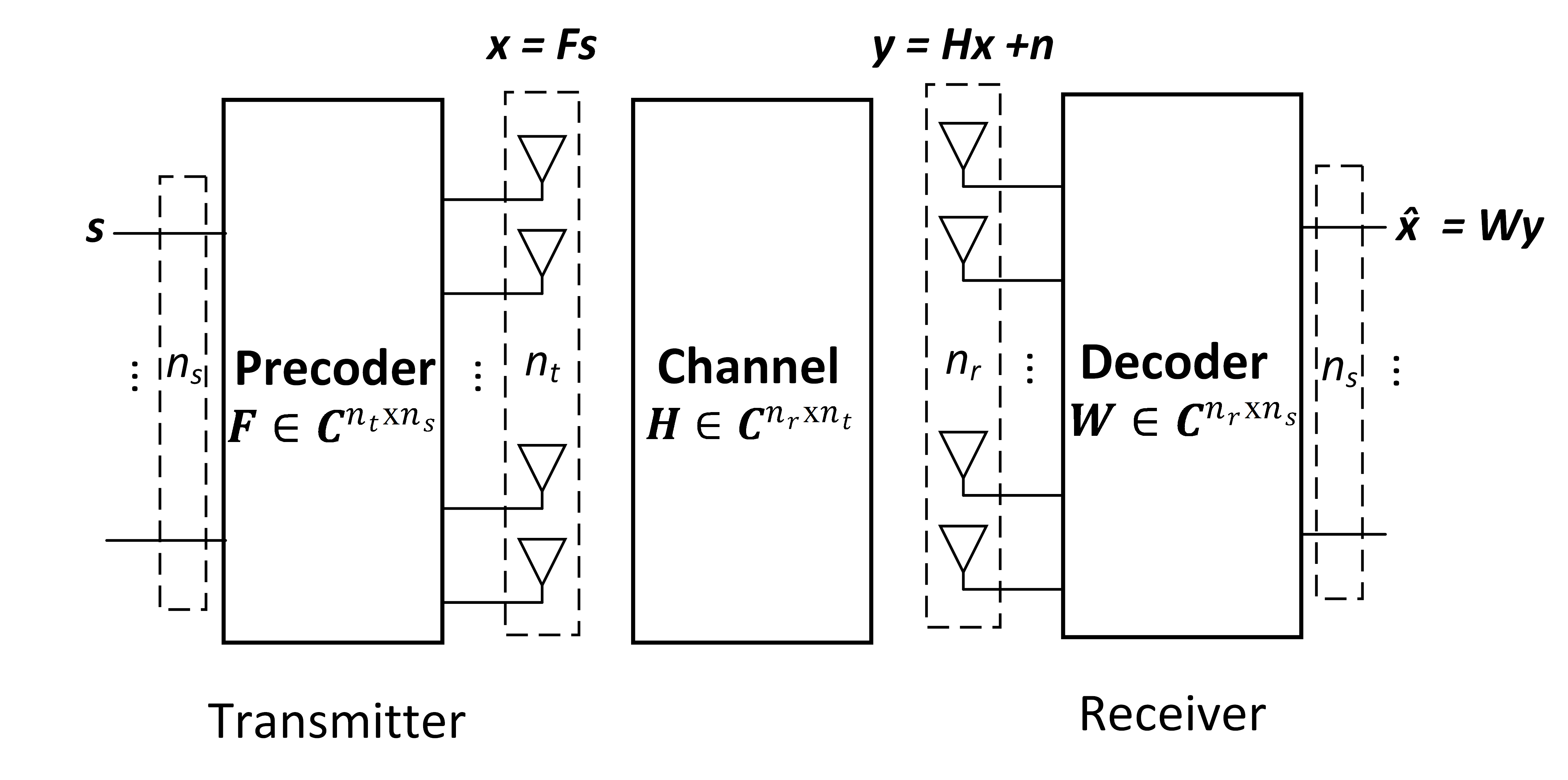}
    \caption{MIMO architecture of typical Wi-Fi systems.}
    \label{fig:MIMO_arch}
\end{figure}

\subsection{MIMO Architecture}
The multiple-input and multiple-output (MIMO) architecture of typical Wi-Fi systems is depicted in Fig. \ref{fig:MIMO_arch}. The transmitted signal and the received signal can be described with the following equation
\begin{equation}
    \begin{aligned}
        \hat{\mathbf{x}}=\mathbf{W}\mathbf{H}\mathbf{F}\mathbf{s}+\mathbf{W}\mathbf{n},
    \end{aligned}
\end{equation}
where $\mathbf{x} \in \mathbf{C}^{n_{r}}$, $\mathbf{W} \in \mathbf{C}^{n_{s} \times n_{r}}$, $\mathbf{H} \in \mathbf{C}^{n_{r} \times n_{t}}$, $\mathbf{F} \in \mathbf{C}^{n_{t} \times n_{s}}$, $\mathbf{s} \in \mathbf{C}^{n_{s}}$, and $\mathbf{n} \in \mathbf{C}^{n_{r}}$  denote the baseband signal at the receiver, decoding matrix, channel matrix, precoding matrix, basedband signal at the transmitter, and the noise vector at the receiving antenna which conforms to zero mean circularly symmetric complex Gaussian (ZMCSCG) with variance $\sigma^{2}$. In our work, we employ the minimum mean squared error (MMSE) detector which is expressed as \cite{sayed2003fundamentals,eraslan2013performance}
\begin{small}\begin{align}\label{eqn: W_perfect}
\mathbf{W}= \left[(\mathbf{H}\mathbf{F})^{*}\mathbf{H}\mathbf{F}+\frac{\mathbf{I}}{\text{SNR}}\right]^{-1}(\mathbf{H}\mathbf{F})^{*},
\end{align}\end{small}
where $\text{SNR} = \frac{\mathbb{E}[(\mathbf{F}\mathbf{s})^{*}(\mathbf{F}\mathbf{s})]}{\mathbb{E}[\mathbf{n}^{*}\mathbf{n}]}$ represents the ratio between the signal energy at each transmit antenna and the noise energy, and $\mathbf{I}$ represents the identity matrix.

\begin{figure}[t]
    \centering
    \includegraphics[width=.18\textwidth]{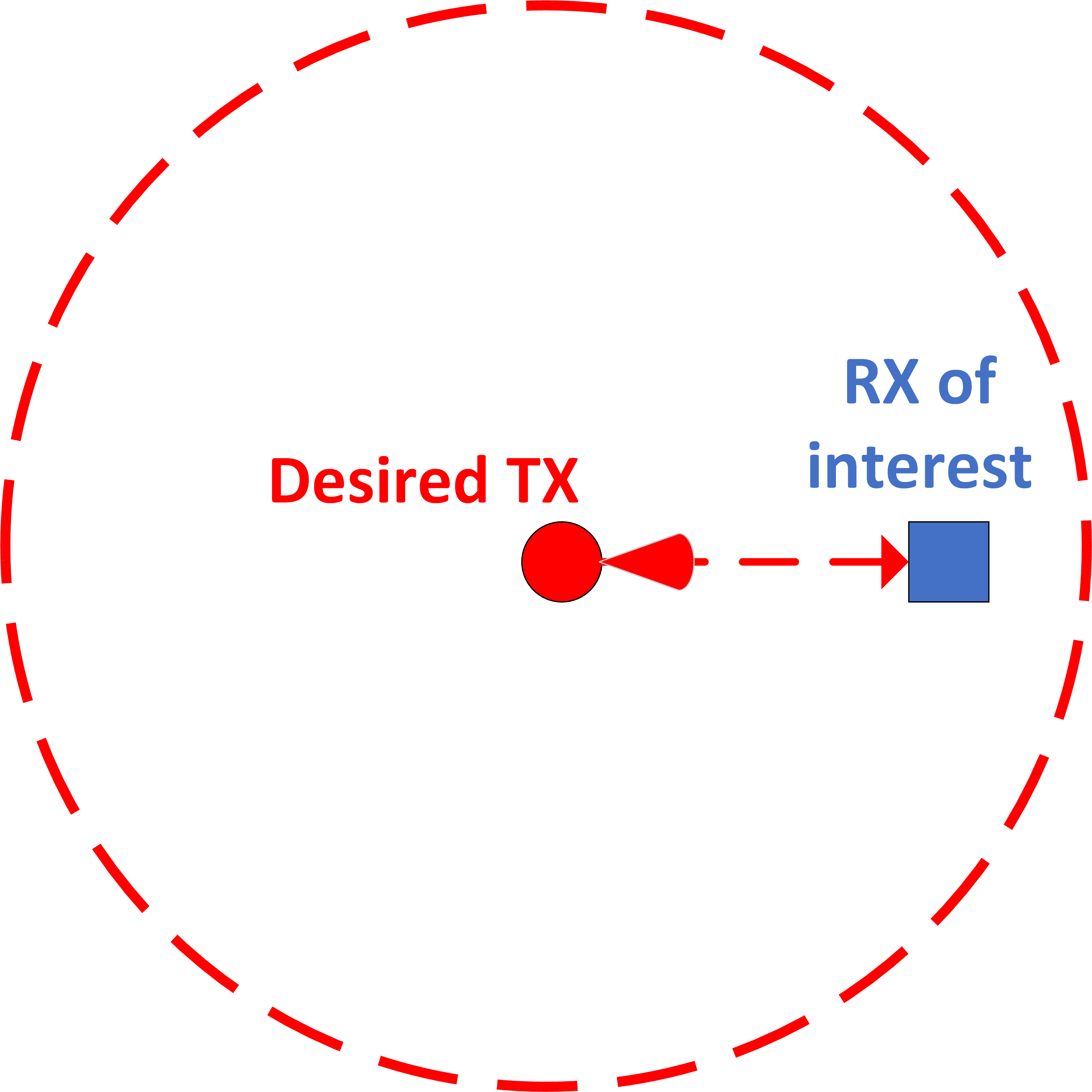}
    \caption{Single BSS single user OFDM scenario.}
    \label{fig:singleBSS}
\end{figure}

As is shown in Fig. \ref{fig:singleBSS}, a desired transmitter transmits $n_{s}$ spatial streams (independent information flows) to a single user using the set of subcarriers $\mathcal N_{sc}$ through Orthogonal Frequency Division Multiplexing (OFDM) in a single basic service set (BSS).
On each subcarrier $i \in \mathcal N_{sc}$, the modulated $n_{s}$ spatial streams are then mapped into $n_t$ transmit antennas using a $n_t \times n_{s}$ precoding matrix $\mathbf{F}_{i}$ for subcarrier $i$~\cite{heath}. The desired transmitted packet is passed through the $n_{r} \times n_{t}$ frequency-domain channel matrices $\mathbf{H}_{i}, i \in \mathcal N_{sc}$ and arrives at the $n_{r}$ receive antennas of receiver.
The frequency-domain channel matrices $\mathbf{H}_{i}$ characterizes the frequency-selectivity property in which $\mathbf{H}_{i}$ varies with subcarrier index $i$ under a frequency-selective channel.

\subsection{Link-to-system (L2S) mapping}

At receiver's $i$-th subcarrier $i \in \mathcal N_{sc}$ and for stream $j \in \{1,2,\ldots, n_{s}\}$, the post processing SINR $\Gamma_{i,j}$ is~\cite{11axEva}
\begin{small}\begin{align}\label{eqn: gamma_mumimo_mmse}
\Gamma_{i,j}= \frac{S_{i,j}}{I^s_{i,j} + N_{i,j}},  
\end{align}\end{small}
where $S_{i,j}$ is the received signal power, $I^s_{i,j}$ is the inter-stream interference, and $N_{i,j}$ is the post processing noise power.
By~\cite{11axEva}, we have $S_{i,j} = P_{t} \left|[\mathbf{W}_{i}]_j^{*} \mathbf{H}_{i} [\mathbf{F}_{i}]_j\right|^2$, $I^s_{i,j} = P_{t}||[\mathbf{W}_{i}]_j^{*} \mathbf{H}_{i} \mathbf{F}_{i}||^2 - S_{i,j}$,  and $N_{i,j} = \sigma^2 ||[\mathbf{W}_{i}]_j||^2$, where 
$P_{t}$ is the received signal power at receiver from the desired transmitter,  $\sigma^2$ is additive noise power on each subcarrier of receiver, $[\cdot]_{j}$ denotes the $j$-th column of a matrix, and $||\cdot||$ is the  Euclidean norm of a vector.
The post processing SINR matrix at receiver is defined by $\boldsymbol \Gamma \triangleq (\Gamma_{i,j})_{i \in \mathcal N_{sc}, 1 \leq j \leq n_{s}}$, which reflects channel frequency-selectivity, antenna correlation captured by $\mathbf{H}_{i}, i \in \mathcal N_{sc}$, and transmit beamforming gain captured by $\mathbf{F}_{i}, i \in \mathcal N_{sc}$ \cite{jin2021efficient}.  

An L2S mapping function $\Phi$ then compresses the post processing SINRs on different subcarriers and streams into an  effective SINR, which would yield the same instantaneous PER if the simulation was run for an AWGN-SISO channel. The effective SINR at receiver is~\cite{RBIR11ax,11axEva,phyAbsMatlab,phyAbsCompar}
\begin{small}\begin{align} \label{eqn: gamma_eff}
    \Gamma_{eff}^{sinr} = \alpha  \Phi^{-1} \left(\frac{1}{n_{sc}}\frac{1}{n_{s}}\sum_{i \in \mathcal N_{sc}}  \sum_{j=1}^{n_{s}} \Phi\left(\frac{\Gamma_{i,j}}{\beta}\right) \right),
\end{align}\end{small}
where $\Phi^{-1}$ is the inverse L2S mapping function, $\mathcal N_{sc}$ is the set of subcarriers, $n_{sc} \triangleq |\mathcal N_{sc}|$ is the number of subcarriers allocated to the receiver, $n_{s}$ is the number of spatial streams sent to the receiver, $\alpha$ and $\beta$ are L2S mapping tuning parameters that depend on PHY layer configurations (channel type, OFDM MIMO setup, MCS and channel coding). The accuracy of the instantaneous PER prediction depends on the L2S mapping function $\Phi$, such as Exponential Effective SINR Mapping (EESM) and Received Bit Information Rate (RBIR) mapping.
In particular, EESM L2S mapping function is used in our work. For EESM L2S mapping, $\alpha = \beta$ and L2S mapping function $\Phi(x) = \exp(-x)$~\cite{phyAbsCompar}.
Then, Eq. \eqref{eqn: gamma_eff} reduces to \cite{11axEva}
\begin{small} \begin{align} \label{eqn: gamma_eff_eesm}
    \Gamma_{eff}^{sinr} = -\beta \ln \left(\frac{1}{n_{sc}}\frac{1}{n_{s}}\sum_{i \in \mathcal N_{sc}}  \sum_{j=1}^{n_{s}} \exp\left(-\frac{\Gamma_{i,j}}{\beta}\right) \right).
\end{align} \end{small}

\section{System Model}
\label{sec:sys_model}

\begin{figure*}[t]
    \centering
    \includegraphics[width=.7\textwidth]{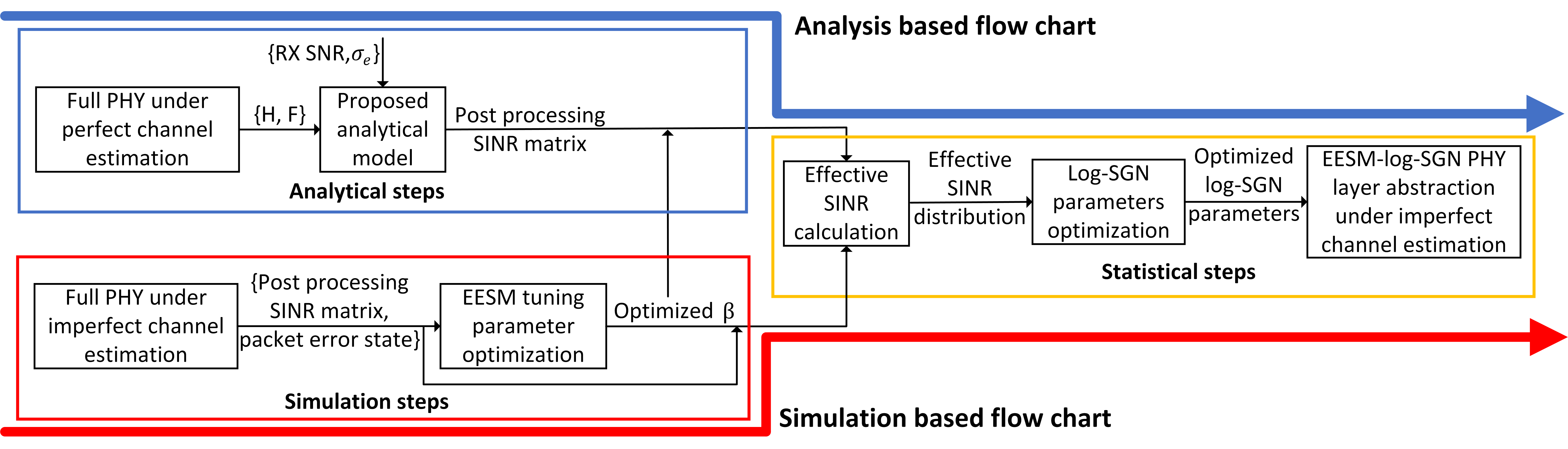}
    \caption{Flow charts for implementing EESM-log-SGN PHY layer abstraction under imperfect channel estimation.}
    \label{fig:FlowChart}
\end{figure*}

\subsection{Simulation based PHY layer abstraction}

The definition of the post processing SINR $\Gamma_{i,j}$ expressed in Eq. (\ref{eqn: gamma_mumimo_mmse}) holds if the channel is perfectly known at the receiver. However, in any real system, the channel matrix at any sub-carrier $\mathbf{H}_i, \forall i \in \mathcal N_{sc}$ estimated by the receiver is inherently noisy, so we model the channel matrix as
\begin{equation}
    \hat{\mathbf{H}}_i = \mathbf{H}_i + \Delta \mathbf{H}_i,
\end{equation}
where $\Delta \mathbf{H}_i$ denotes the estimation error matrix assumed to be uncorrelated with $\mathbf{H}_i$, and each entry of $\mathbf{H}_i$ follows ZMCSCG with variance $\sigma_e^2$ \footnote{The channel estimation noise variance can be defined as $\sigma_e^2 = \frac{1}{n_{t}SNR}$ \cite{hassibi2003much}.}, The quality of channel estimation is captured by $\sigma_e^2$, which can be appropriately estimated depending on the channel estimation method. We assume that each block (packet), that undergoes a different channel realization $\mathbf{H}_i$, observes a different realization of $\Delta \mathbf{H}_i$ at the receiver. As a result, the packet-level performance may be significantly impacted by the quality of the channel estimation. 
\begin{figure}[h]
    \centering
    \includegraphics[width=.35\textwidth]{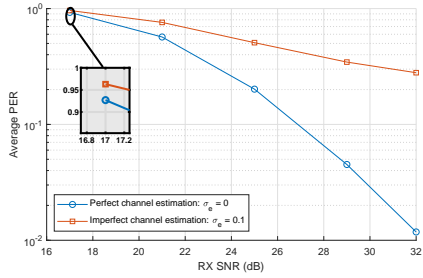}
    \caption{Full PHY simulation: Average PER versus SNR under OFDM allocation with 242 subcarriers, $2\times2$ MIMO,
IEEE TGax channel model-B, MCS5.}
    \label{fig:pervssnr}
\end{figure}

As is shown in Fig. \ref{fig:FlowChart}, we first demonstrate the simulation-based flow chart that consists of the simulation steps and statistical steps. In the simulation steps, we develop the full PHY simulation \footnote{The full PHY simulation is conducted using a credible link simulator (e.g., MATLAB WLAN Toolbox) to provide calibrated L2S mapping tuning parameters.} considering imperfect channel estimation. Based on the developed the full PHY simulation, we can obtain the packet-level performance under imperfect channel estimation. Fig. \ref{fig:pervssnr} shows the average PER regarding the SNR under perfect and imperfect channel estimation through the full PHY simulation. One can observe that a small $\sigma_e$ can cause large PER difference between different channel estimation qualities in each SNR points. 

Subsequently, the post processing SINR matrix and the associated binary packet error state \footnote{The frequency-selective channel instance is assumed to be almost invariant during the transmission of each packet (e.g., IEEE TGax channel models \cite{11axChannel}) and different channel instances for different packets are assumed to be i.i.d.~\cite{phyAbsMatlab,perEstimationMatlab,perEstimationMatlab2}.} for each packet can be obtained after the full PHY simulation. The set \{post processing SINR matrix $\boldsymbol \Gamma$, binary packet error state\} are then used to optimize the EESM tuning parameter $\beta$ of the EESM mapping function \cite{cipriano2008calibration}. The parameter $\beta$ in Eq. (\ref{eqn: gamma_eff}) is then optimized to minimize the mean square error (MSE) between the instantaneous PER-effective SINR curve for the simulated frequency-selective fading channel and the instantaneous PER-SNR curve under the AWGN-SISO channel. Then, in the initial stage of the statistical steps, an effective SINR distribution can be obtained through Eq. (\ref{eqn: gamma_eff}) with the optimized EESM parameter $\beta$ and post processing SINR matrices of all packets. 

\begin{lemma}
\label{lemma:overlap}
The effective SINR $\Gamma_{eff}^{sinr}$ is irrelevant to the channel estimation error $\sigma_e$ under MISO/SISO configuration.
\begin{proof}
Under MISO configuration, $\mathbf{W}_{i}$ becomes a scalar, i.e., $W_{i}$, as $n_{r}=1$ and $ n_{s}=1$. For the $i$-th subcarrier $i \in \mathcal N_{sc}$ and the only stream (indicating $I^s_{i,1} = 0$), the post processing SINR $\Gamma_{i,1}$ is expressed as $\Gamma_{i,1}= \frac{S_{i,1}}{N_{i,1}}$, where $S_{i,1} = P_{t} \left|[W_i]_1^{*} \mathbf{H}_{i} [\mathbf{F}_{i}]_1\right|^2 = P_{t} W_{i}^2 \left| \mathbf{H}_{i} [\mathbf{F}_{i}]_1\right|^2$, and $N_{i,1} = \sigma^2 ||[W_{i}]_1||^2 = \sigma^2W_{i}^2$. Substitute  $S_{i,1}$ and $N_{i,1}$ with the above extended terms, we get $\Gamma_{i,1}= \frac{P_{t} \left| \mathbf{H}_{i} [\mathbf{F}_{i}]_1\right|^2}{\sigma^2}$. Similarly, under SISO configuration, $\mathbf{W}_{i}$, $\mathbf{H}_{i}$ and $\mathbf{F}_{i}$ are all scalars, i.e., $W_{i}$, $H_{i}$ and $F_{i}$,  as $n_{r}=1$, $ n_{s}=1$ and $n_{t} = 1$. Hence $S_{i,1} = P_{t} \left|[W_{i}]_1^{*} H_{i} F_{i}\right|^2 = P_{t} W_{i}^2 H_{i}^2 F_{i}^2$ and $N_{i,1} = \sigma^2 || [W_{i}]_1 ||^2 = \sigma^2 W_{i}^2$. Then we get $\Gamma_{i,1}= \frac{P_{t} H_{i}^2 F_{i}^2}{\sigma^2}$. Hence based on Eq. (\ref{eqn: gamma_eff_eesm}), the effective SINR under both configurations becomes $\Gamma_{eff}^{sinr} = -\beta \Phi^{-1} \left(\frac{1}{n_{sc}}\sum_{i \in \mathcal N_{sc}} \Phi\left(\frac{\Gamma_{i,1}}{\beta}\right) \right)$, which is irrelevant to $W_{i}$ impacted by $\sigma_e$.
\end{proof}
\end{lemma}

\begin{remark}
According to Lemma \ref{lemma:overlap}, the channel estimation error does not impact effective SINR distribution under MISO/SISO configuration. However, this does not hold under MIMO/SIMO configuration as the MMSE equalizer, $\mathbf{W}_{i}$, cannot be cancelled when calculating the post processing SINR.
\end{remark}

\begin{figure}[t]
    \centering
    \includegraphics[width=.3\textwidth]{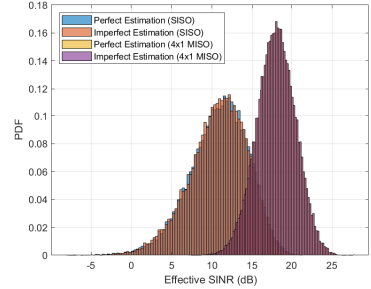}
    \caption{Effective SINR distributions following simulation steps (SNR = 13dB).}
    \label{fig:sisoMisoOverlap}
\end{figure}
As Fig.\ref{fig:sisoMisoOverlap} shows, the effective SNR distributions following the simulation steps overlap in the MISO/SISO case under the perfect and imperfect channel estimation. In the statistical steps, the obtained effective SINR distribution is then statistically fitted with the log-SGN distribution \cite{jin2021efficient} which is defined as
\begin{small} \begin{align}\label{eqn: XSGN}
    X \triangleq \ln (\Gamma_{eff}^{sinr}) \sim {\rm SGN} (\hat{\mu}, \hat{\sigma}, \hat{\lambda}_1, \hat{\lambda}_2),
\end{align}\end{small}
with PDF
\begin{small} \begin{align}
    &f_X(x;\hat{\mu}, \hat{\sigma}, \hat{\lambda}_1, \hat{\lambda}_2) \nonumber \\
    =& \frac{2}{\hat{\sigma}}  \psi \left(\frac{x-\hat{\mu}}{\hat{\sigma}}\right) \Psi \left(\frac{\hat{\lambda}_1(x - \hat{\mu})}{\sqrt{\hat{\sigma}^2 + \hat{\lambda}_2(x - \hat{\mu})^2}}\right),  
     \ x \in \mathbf{R},
\end{align}\end{small}
where $\hat{\mu} \in \mathbf{R}$ is the location parameter, $\hat{\sigma}>0$ is the scale  parameter, $\hat{\lambda}_1 \in \mathbf{R}$ and  $\hat{\lambda}_2 \geq 0$ are shape parameters, $\psi(x)$ is the standard normal PDF, and $\Psi(x)$ is the standard normal cumulative distribution function. As the effective SINRs for log-SGN fitting are based on full PHY simulation results under imperfect channel estimation, the four log-SGN parameters summarize the key information in the PHY layer in the existence of channel estimation error.

\subsection{Analysis based PHY layer abstraction}
Fig. \ref{fig:FlowChart} shows an analysis based flow chart to achieve the EESM-log-SGN PHY layer abstraction under imperfect channel estimation. This flow chart includes the analytical steps as well as the statistical steps. In the analytical steps, we use the full PHY simulation under perfect channel estimation to track the (perfect) channel realization information (i.e., channel matrix $\mathbf{H}_i, i \in \mathcal N_{sc}$ and precoding matrix $\mathbf{F}_i, i \in \mathcal N_{sc}$ of each packet). Then an analytical model is developed to characterize the relation between the channel estimation error and the effective SINR. The developed analytical model relies on the SNR, channel estimation error $\sigma_e^2$ and the information from the full PHY simulation under perfect channel estimation. After the analytical steps, the analytical post processing SINRs under imperfect channel estimation are obtained. Then we can develop the EESM-log-SGN PHY abstraction considering channel estimation error based on the analytical effective SINR distribution which is processed by the statistical steps. Note that given a fixed SNR, different packets characterize different $\hat{\Gamma}^{sinr}_{eff}$ due to the block fading channels. Thus, the analytical steps output a distribution rather than a number. In the rest of this section, we present the analytical model of effective SINR for MMSE detector with channel estimation error.
\begin{figure*}[t]
\begin{small} \begin{equation}\label{eq:long1}
    \mathbf{E}_{i} = \begin{pmatrix} \sigma_{e}^{2} \text{tr}(\mathbf{K}_i\mathbf{F}_i^{*}\mathbf{H}_i^{*}\mathbf{H}_i\mathbf{F}_i\mathbf{F}_i^{*}\mathbf{H}_i^{*}\mathbf{H}_i\mathbf{F}_i\mathbf{K}^{*}_i)\mathbf{K}_i\mathbf{F}_i^{*}\mathbf{H}_i^{*}\mathbf{H}_i\mathbf{F}_i\mathbf{K}_i^{*}+\sigma_{e}^{2}\text{tr}(\mathbf{H}_i\mathbf{F}_i\mathbf{K}_i\mathbf{F}_i^{*}\mathbf{H}_i^{*}\mathbf{H}_i\mathbf{F}_i\mathbf{F}_i^{*}\mathbf{H}_i^{*}\mathbf{H}_i\mathbf{F}_i\mathbf{K}^{*}_i\mathbf{F}_i^{*}\mathbf{H}_i^{*})\mathbf{K}_i\mathbf{K}^{*}_i\\-\sigma_{e}^{2}\text{tr}(\mathbf{H}_i\mathbf{F}_i\mathbf{K}_i\mathbf{F}_i^{*}\mathbf{H}_i^{*}\mathbf{H}_i\mathbf{F}_i\mathbf{F}_i^{*}\mathbf{H}_i^{*})\mathbf{K}_i\mathbf{K}^{*}_i  -\sigma_{e}^{2}\text{tr}(\mathbf{H}_i\mathbf{F}_i\mathbf{F}_i^{*}\mathbf{H}_i^{*}\mathbf{H}_i\mathbf{F}_i\mathbf{K}^{*}_i\mathbf{F}_i^{*}\mathbf{H}_i^{*})\mathbf{K}_i\mathbf{K}^{*}_i\\+\sigma_{e}^{2}\text{tr}(\mathbf{H}_i\mathbf{F}_i\mathbf{F}_i^{*}\mathbf{H}_i^{*})\mathbf{K}_i\mathbf{K}^{*}_i+\text{SNR}^{-1}\mathbf{W}_i\mathbf{W}_i^{*} \end{pmatrix}
\end{equation} \end{small}
\end{figure*}

\begin{figure*}[t]
\begin{small}
\begin{equation}\label{eq:long2}
     \mathbf{N}_i = \begin{pmatrix} \text{SNR}^{-1}\mathbf{W}_i\mathbf{W}_i^{*}+\text{SNR}^{-1}\sigma_{e}^{2}\text{tr}(\mathbf{K}_i\mathbf{F}_i^{*}\mathbf{H}_i^{*}\mathbf{H}_i\mathbf{F}_i\mathbf{K}^{*}_i)\mathbf{K}_i\mathbf{F}_i^{*}\mathbf{H}_i^{*}\mathbf{H}_i\mathbf{F}_i\mathbf{K}^{*}_i\\+\text{SNR}^{-1}\sigma_{e}^{2}\text{tr}(\mathbf{H}_i\mathbf{F}_i\mathbf{K}_i\mathbf{F}_i^{*}\mathbf{H}_i^{*}\mathbf{H}_i\mathbf{F}_i\mathbf{K}^{*}_i\mathbf{F}_i^{*}\mathbf{H}_i^{*})\mathbf{K}_i\mathbf{K}^{*}_i-\text{SNR}^{-1}\sigma_{e}^{2}\text{tr}(\mathbf{H}_i\mathbf{F}_i\mathbf{K}_i\mathbf{F}_i^{*}\mathbf{H}_i^{*})\mathbf{K}_i\mathbf{K}^{*}_i\\-\text{SNR}^{-1}\sigma_{e}^{2}\text{tr}(\mathbf{H}_i\mathbf{F}_i\mathbf{K}^{*}_i\mathbf{F}_i^{*}\mathbf{H}_i^{*})\mathbf{K}_i\mathbf{K}^{*}_i+\text{SNR}^{-1}\sigma_{e}^{2}N_{r}\mathbf{K}_i\mathbf{K}^{*}_i \end{pmatrix}
\end{equation}\end{small}
\end{figure*}

Following Eq. (\ref{eqn: W_perfect}), the receiver can use the imperfect estimated channel to calculate the MMSE detector as follows,
\begin{small}
\begin{equation}
    \begin{aligned}
        \hat{\mathbf{W}}_i &=[(\mathbf{H}_i\mathbf{F}_i+\Delta\mathbf{H}_i\mathbf{F}_i)^{*}(\mathbf{H}_i\mathbf{F}_i+\Delta\mathbf{H}_i\mathbf{F}_i)+\frac{\mathbf{I}}{\text{SNR}}]^{-1}\bullet\\&~~~~~~~~~~~~~~~~~~~~~~~~~~~~~~~~~~~~~~~~~~~~~~(\mathbf{H}_i\mathbf{F}_i+\Delta\mathbf{H}_i\mathbf{F}_i)^{*}\\&= [\mathbf{F}^{*}_i\mathbf{H}^{*}_i\mathbf{H}_i\mathbf{F}_i+\frac{\mathbf{I}}{\text{SNR}}+\mathbf{F}^{*}_i\mathbf{H}^{*}_i\Delta\mathbf{H}_i\mathbf{F}_i\\&~~~~~~~~~~~~~~~~~~~~+\mathbf{F}^{*}_i\Delta\mathbf{H}^{*}_i\mathbf{H}_i\mathbf{F}_i]^{-1}\bullet \mathbf{F}^{*}_i(\mathbf{H}_i+\Delta\mathbf{H}_i)^{*},
    \end{aligned}
\end{equation}
\end{small}where we ignore the term $\mathbf{F}^{*}_i\Delta\mathbf{H}^{*}_i\Delta\mathbf{H}_i\mathbf{F}_i$ since it is much less significant than the channel estimation error $\Delta\mathbf{H}_i$ when $\sigma_{e}^{2}$ is small. With imperfect channel estimation, $\Delta \mathbf{H}_i$, the imperfect MMSE solution can also be written as $\hat{\mathbf{W}}_i = \mathbf{W}_i
+\Delta \mathbf{W}_i$. Thus the MMSE estimate of the signal vector becomes 
\begin{small}\begin{equation}
    \hat{\mathbf{x}}_i = \hat{\mathbf{W}}_i\mathbf{y}_i=\mathbf{W}_i\mathbf{H}_i\mathbf{x}_i+\Delta\mathbf{W}_i\mathbf{H}_i\mathbf{x}_i+\mathbf{W}_i\mathbf{n}_i+\Delta\mathbf{W}_i\mathbf{n}_i,
\end{equation}\end{small}
where $\mathbf{x} = \mathbf{F}\mathbf{s}$ represents the precoded base-band data streams, and $\mathbf{s}$ is the original signal. Denote $\mathbf{K}_i=(\mathbf{H}^{*}_i\mathbf{F}^{*}_i\mathbf{F}_i\mathbf{H}_i+\frac{\mathbf{I}}{\text{SNR}})^{-1}$, then based on $(\mathbf{P}+\epsilon^{2}\mathbf{Q})^{-1} \approx \mathbf{P}^{-1} - \epsilon^{2}\mathbf{P}^{-1}\mathbf{Q}\mathbf{P}^{-1}$ given small $\epsilon^{2}$ we have the following,
\begin{small}\begin{equation}
\begin{aligned}
    \hat{\mathbf{W}}_i &= (\mathbf{K}_i-\mathbf{K}_i(\mathbf{F}^{*}_i\mathbf{H}^{*}_i\Delta\mathbf{H}_i\mathbf{F}_i+\mathbf{F}^{*}_i\Delta\mathbf{H}^{*}_i\mathbf{H}_i\mathbf{F}_i)\mathbf{K}_i)\bullet\\&~~~~~~~~~~~~~~~~~~~~~~~~~~~~~~~~~~~~~(\mathbf{H}_i\mathbf{F}_i+\Delta\mathbf{H}_i\mathbf{F}_i)^{*},
    \end{aligned}
\end{equation}\end{small}
where $\triangle\mathbf{W}_i$ can be approximated as $-\mathbf{K}_i(\mathbf{F}^{*}_i\mathbf{H}^{*}_i\Delta\mathbf{H}_i\mathbf{F}_i+\mathbf{F}^{*}_i\Delta\mathbf{H}^{*}_i\mathbf{H}_i\mathbf{F}_i)\mathbf{K}_i\mathbf{F}^{*}_i\mathbf{H}^{*}_i+\mathbf{K}_i\mathbf{F}^{*}_i\Delta\mathbf{H}^{*}_i$.
\begin{small}\begin{equation}
\begin{aligned}
    &\mathbb{E}[(\hat{\mathbf{W}}_i\mathbf{H}_i\mathbf{F}_i)(\hat{\mathbf{W}}_i\mathbf{H}_i\mathbf{F}_i)^{*}] \\=&\mathbb{E}[(( \mathbf{W}_{i}+\Delta \mathbf{W}_{i})\mathbf{H}_i\mathbf{F}_i)((\mathbf{W}_{i}+\Delta \mathbf{W}_{i})\mathbf{H}_i\mathbf{F}_i)^{*}] \\=& \mathbb{E}[\mathbf{W}_{i}\mathbf{H}_i\mathbf{F}_i\mathbf{F}_i^{*}\mathbf{H}_i^{*}\mathbf{W}_i^{*}] +\mathbb{E}[\Delta\mathbf{W}_{i}\mathbf{H}_i\mathbf{F}_i\mathbf{F}_i^{*}\mathbf{H}_i^{*}\Delta\mathbf{W}_i^{*}] \\=& \mathbf{W}_{i}\mathbf{H}_i\mathbf{F}_i\mathbf{F}_i^{*}\mathbf{H}_i^{*}\mathbf{W}_i^{*} + \mathbf{E}_{i}
\end{aligned}
\end{equation}\end{small}
where $\mathbf{E}_{i}$ is obtained with property $\mathbb{E}[\Delta\mathbf{H}\mathbf{A}\Delta\mathbf{H}]=\mathbb{E}[\Delta\mathbf{H}^{*}\mathbf{A}\Delta\mathbf{H}^{*}] = 0$ and $\mathbb{E}[\Delta\mathbf{H}\mathbf{A}\Delta\mathbf{H}^{*}] = \sigma_{e}^{2} \text{tr}(\mathbf{A})\mathbf{I}$ in Eq. \eqref{eq:long1}. The corresponding SINR is expressed as
\begin{small}
\begin{equation}\label{eq:postmimosinr_analy}
\begin{split}
\begin{aligned}
    &\hat{\Gamma}_{i,j} = \frac{|(\mathbf{W}_{i}\mathbf{H}_i\mathbf{F}_i\mathbf{F}_i^{*}\mathbf{H}_i^{*}\mathbf{W}_i^{*} + \mathbf{E}_{i})_{j,j}|^2}{\sum_{l \neq j}|(\mathbf{W}_{i}\mathbf{H}_i\mathbf{F}_i\mathbf{F}_i^{*}\mathbf{H}_i^{*}\mathbf{W}_i^{*}+\mathbf{E}_{i})_{j,l}|^2 + [\mathbf{N}_i]_{j,j}},
\end{aligned}
\end{split}
\end{equation}\end{small}where
\begin{small}
\begin{equation}
    \begin{aligned}
            &\mathbf{N}_i=\mathbb{E}[\mathbf{W}_i\mathbf{n}_i\mathbf{n}^{*}_i\mathbf{W}^{*}_i]+\mathbb{E}[\mathbf{W}_i\mathbf{n}_i\mathbf{n}^{*}_i\Delta\mathbf{W}^{*}_i]+\mathbb{E}[\Delta\mathbf{W}_i\mathbf{n}_i\mathbf{n}^{*}_i\mathbf{W}^{*}_i]\\&+\mathbb{E}[\Delta\mathbf{W}_i\mathbf{n}_i\mathbf{n}_i\Delta\mathbf{W}^{*}_i]
    \end{aligned}
\end{equation}
\end{small}represents the expected noise power that can be further expressed in Eq. \eqref{eq:long2}, using the property that $\mathbb{E}[\Delta\mathbf{W}] = 0$, $\mathbb{E}[\Delta\mathbf{H}\mathbf{A}\Delta\mathbf{H}]=\mathbb{E}[\Delta\mathbf{H}^{*}\mathbf{A}\Delta\mathbf{H}^{*}]=0$ and $\mathbb{E}[\Delta\mathbf{H}\mathbf{A}\Delta\mathbf{H}^{*}]=\sigma_{e}^{2} \text{tr}(\mathbf{A})\mathbf{I}$. It is noteworthy that the channel matrix $\mathbf{H}_i$ and $\mathbf{F}_i$ in Eq. (\ref{eq:postmimosinr_analy}) are the ground truth matrices which are irrelevant to $\sigma_e^2$. This indicates that the MMSE detector $\mathbf{W}_i$ in Eq. (\ref{eq:postmimosinr_analy}) is also the ground truth matrix as it follows Eq. (\ref{eqn: W_perfect}). Afterwards, the analytical effective SINR under imperfect channel estimation is given by 
\begin{small}\begin{align} 
    \hat{\Gamma}^{sinr}_{eff} = -  \hat{\beta}  \ln \left(\frac{1}{n_{sc}}\frac{1}{n_{s}}\sum_{i \in \mathcal N_{sc}}  \sum_{j=1}^{n_{s}} \exp\left(-\frac{\hat{\Gamma}_{i,j}}{\hat{\beta}}\right) \right),
\end{align}\end{small}
where $\hat{\Gamma}_{i,j}$ is expressed in Eq. (\ref{eq:postmimosinr_analy}), and $\hat{\beta}$ is obtained from the simulation steps. Finally, the obtained effective SINR distribution under imperfect channel estimation is processed through the statistical steps, where the fitted log-SGN distribution follows $\hat{X} \triangleq \ln (\hat{\Gamma}_{eff}^{sinr})$ expressed in Eq. (\ref{eqn: XSGN}).

\section{Validation}
\label{sec:sim}

\begin{table}[t]
 \centering
 \caption{\small{PHY layer simulation setup.} 
}\label{tab: phy_setup}
\resizebox{.4\textwidth}{!}{\begin{tabular}{ |c|c|c|c|c|c|c| } 
\hline
Communication system & IEEE 802.11ax \\
\hline
Link simulator & MATLAB WLAN Toolbox R2021b\\
\hline
Number of packets/simulation & 20000 \\
\hline
Channel type  & IEEE TGax channel model-B~\cite{11axChannel} \\
\hline
Channel for each packet & i.i.d. \\
\hline
Speed of the scatter/user & $0.089$km/h \\
\hline
Channel coding & LDPC \\
\hline
Payload length & 1000\\
\hline
MCS & 5 \\
\hline
Bandwidth & $20$ MHz ($242$ subcarriers in total) \\
\hline
Channel estimation error & ZMCSCG \\
\hline
MIMO/MU-MIMO PHY & SVD/ZF precoding, MMSE decoding \\
\hline
CPU & Intel Core i7 CPU at 2.60GHz \\
\hline
\end{tabular}}
\end{table}

We validate effective SINRs following the simulation based flow chart through the ones from the analysis based flow chart under different OFDM scenarios. The main PHY layer simulation setup is summarized in Table \ref{tab: phy_setup}.


\begin{figure}[t]
    \centering
    \includegraphics[width=.3\textwidth]{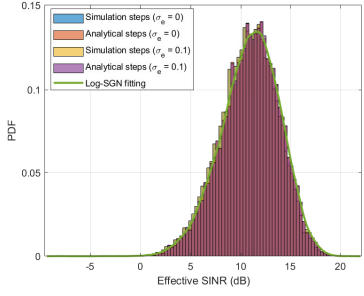}
    \caption{Effective SINR distribution: Perfect/Imperfect channel estimation under $2\times1$ MISO (SNR = 9dB) configuration.}
    \label{fig:miso2x1imperfect}
\end{figure}

We first validate the MISO configuration shown in Fig. \ref{fig:miso2x1imperfect}. The effective SINR distribution obtained from the analytical steps matches well with the one obtained from the simulation steps. Meanwhile, the effective SINR distributions overlap under perfect/imperfect channel estimation, which coincides with Lemma \ref{lemma:overlap}. Additionally, we use the log-SGN fitting method to get the log-SGN parameters characterized by Eq. (\ref{eqn: XSGN}). As the log-SGN fitting curve matches the analytical SINR distribution, the obtained log-SGN parameters accurately describe the system level performance under PHY layer configuration. 

\begin{figure}[ht]
\begin{minipage}[t]{0.48\linewidth}
\centering
 \subfigure[SIMO with $\sigma_e = 0$.]
{\includegraphics[width=1.7in]{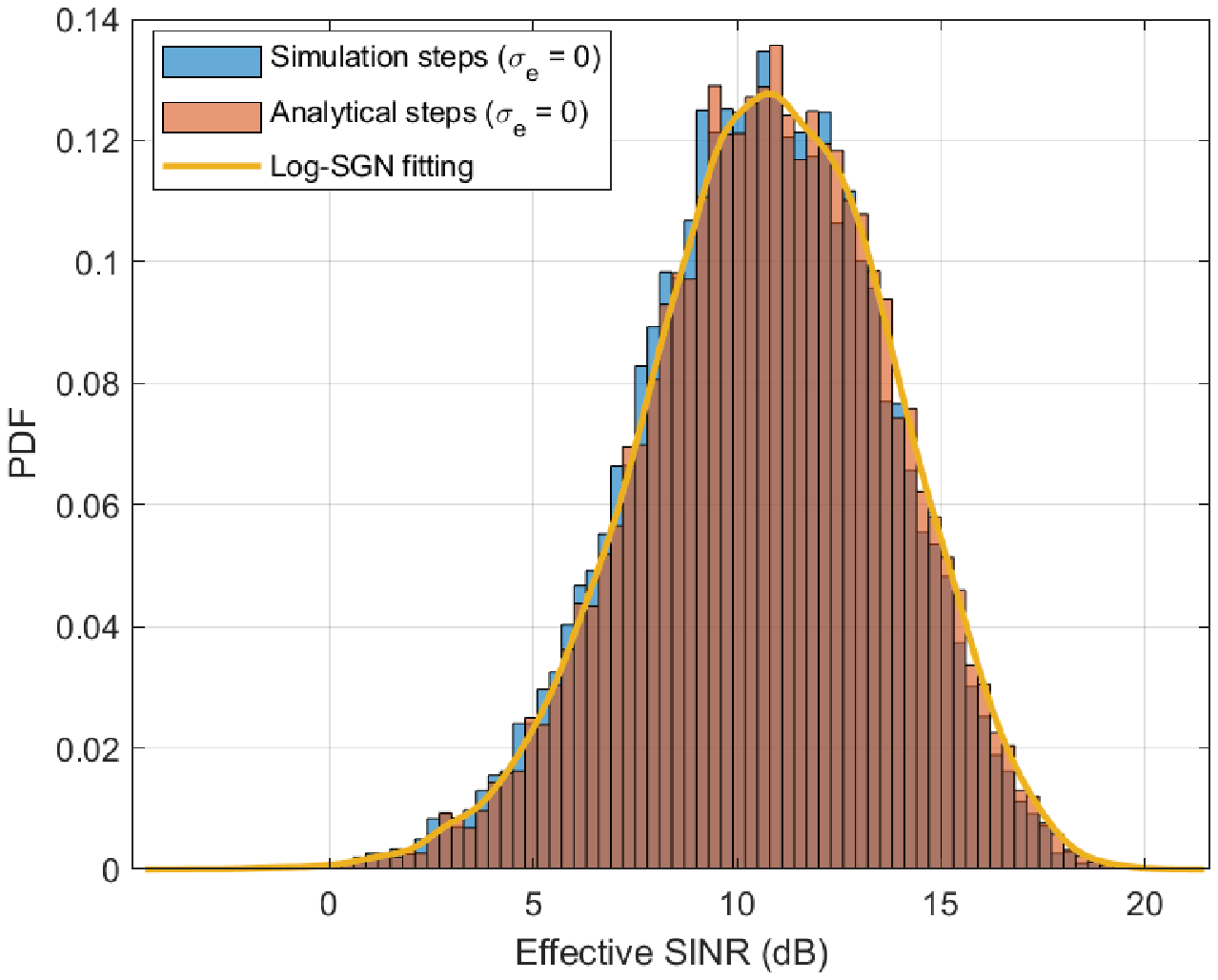}}
\label{fig:7a}
\end{minipage}
\begin{minipage}[t]{0.24\linewidth}
\centering
\subfigure[SIMO with $\sigma_e = 0.1$.]{
\includegraphics[width=1.7in]{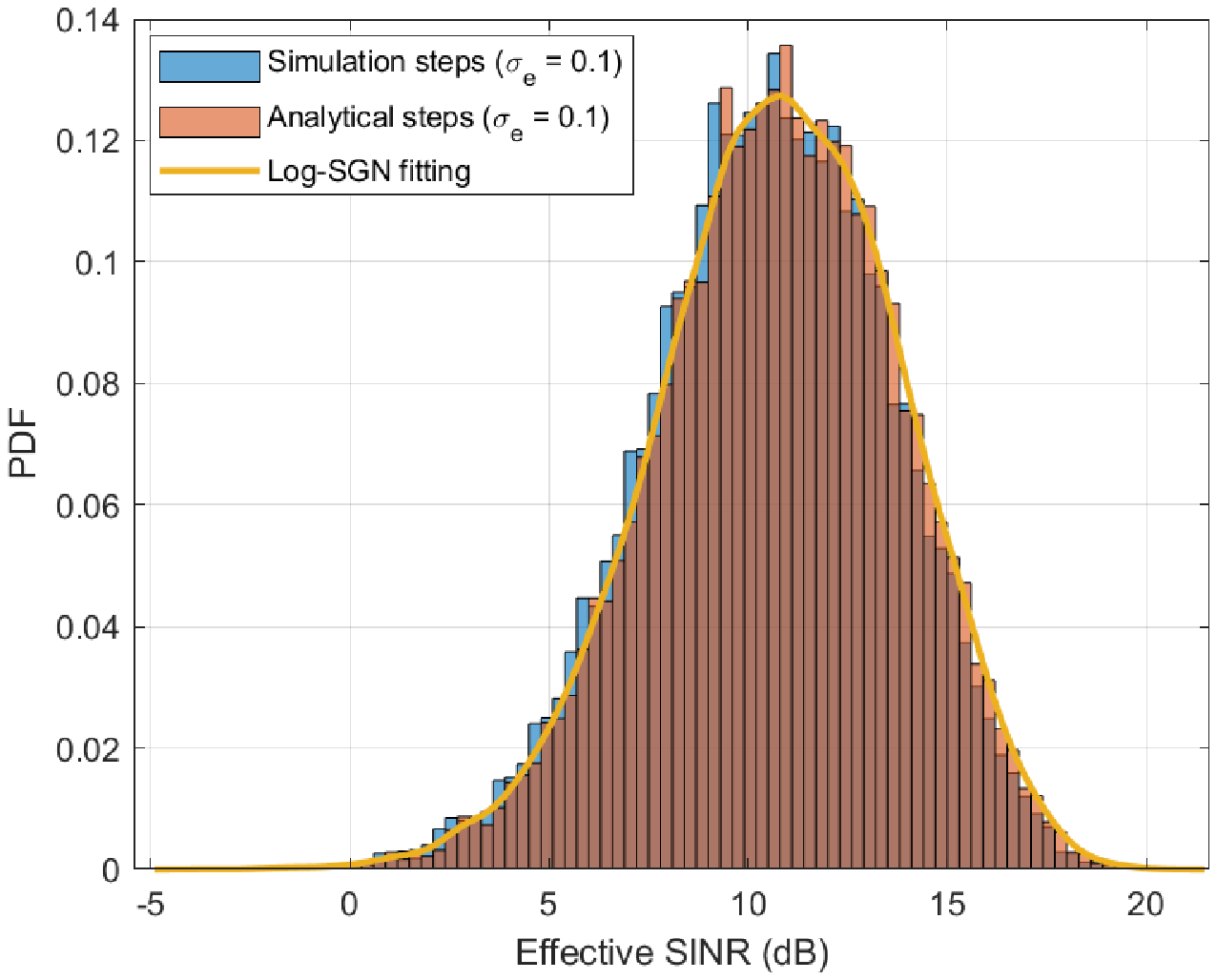}}
\label{fig:7b}
\end{minipage}%

\medskip

\begin{minipage}[t]{0.48\linewidth}
\centering
\subfigure[MIMO with $\sigma_e = 0$.]{
\includegraphics[width=1.7in]{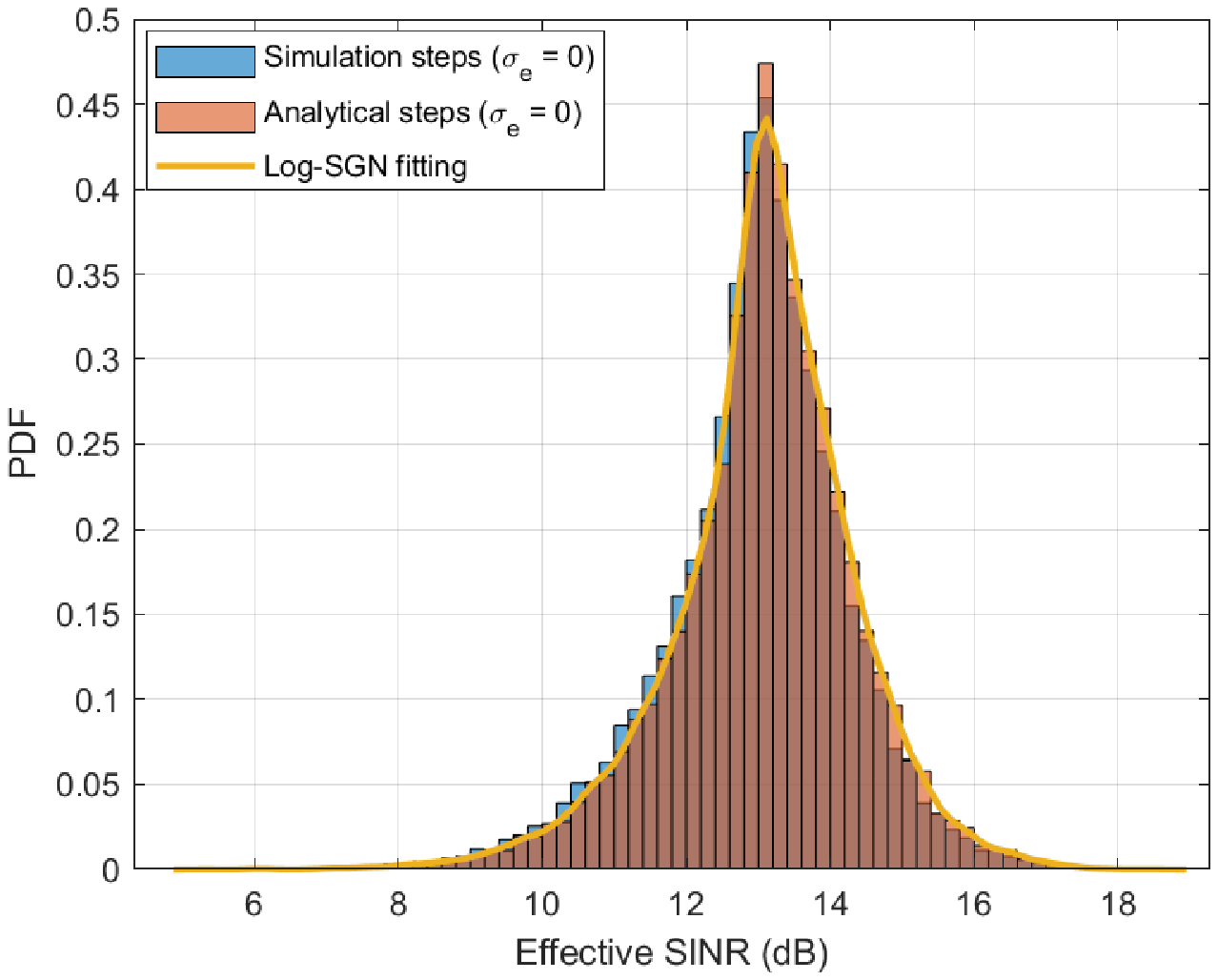}}
\label{fig:7c}
\end{minipage}
\begin{minipage}[t]{0.24\linewidth}
\centering
\subfigure[MIMO with $\sigma_e = 0.1$.]{
\includegraphics[width=1.7in]{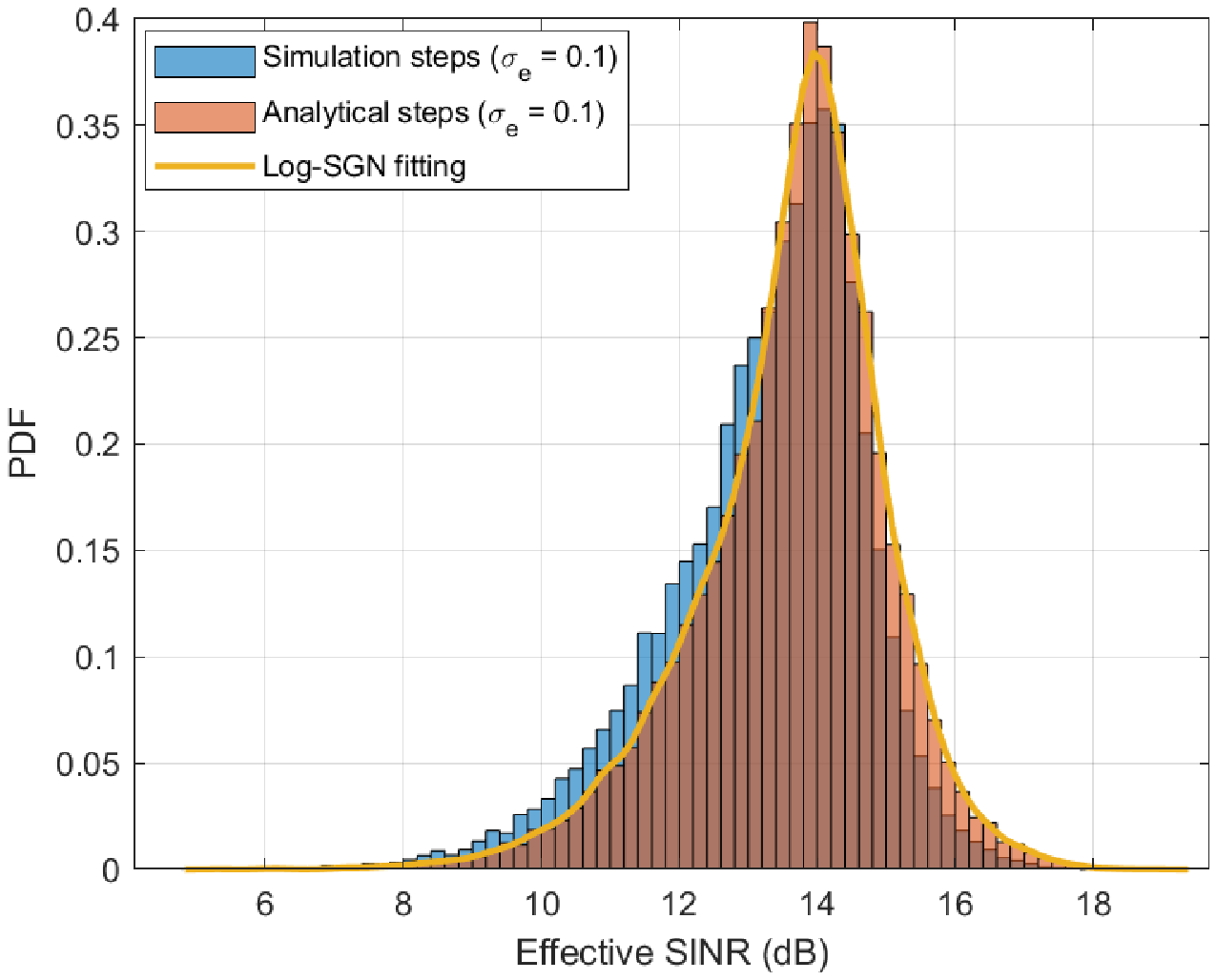}}
\label{fig:7d}
\end{minipage}
\caption{Effective SINR distribution: Perfect/Imperfect channel estimation under $1\times2$ SIMO (SNR = 9dB)/$2\times2$ MIMO (SNR = 17dB) configuration.}
\label{fig:simoMiso}
\end{figure}

Moreover, we validate the developed methods under single-input and multiple-output (SIMO)/MIMO configuration. As Fig. \ref{fig:simoMiso} shows, the analytical effective SINR distribution fits with the corresponding the one obtained from the simulation steps under both configurations. Notice that compared with the SIMO configuration (Fig. \ref{fig:simoMiso}(a) and (b)), the shapes of effective SINR distribution under MIMO configuration (Fig. \ref{fig:simoMiso}(c) and (d)) are more divergent with the same setup of channel estimation quality. This is because the channel estimation error in under MIMO configuration impacts the post processing SINR more significantly than that under SIMO configuration. Similar to the MISO configuration, the log-SGN fitting curves also accurately characterizes the analytical effective SINR distributions under both configurations. Therefore, the log-SGN distribution can describe the effective SINR under all configurations considering the channel estimation error.

\section{Conclusion}
\label{sec:con}
In this letter, we investigated the EESM-log-SGN PHY layer abstraction with the channel estimation error. Two methods were developed to obtain the target PHY layer abstraction. The effective SINR was demonstrated to be irrelevant to the channel estimation error under MISO/SISO configuration. We validated that the effective SINRs from the analytical steps fitted with the ones from the simulation steps. Meanwhile, the log-SGN distribution is verified to accurately describe effective SINRs under imperfect channel estimation.

\bibliographystyle{IEEEtran}

\bibliography{ref}
\end{document}